\documentclass[]{raa}
\usepackage{amssymb}            
\usepackage{graphicx,times}
\usepackage{cases}

\providecommand{\bm}[1]{\mbox{\boldmath$#1$\unboldmath}}

\begin{document}

   \title{The Potential of FAST in Detecting Celestial Hydroxyl Masers and Related Science Topics$^*$
\footnotetext{\small $*$ Supported by China Ministry of Science and Technology under State Key Development Program for Basic Research (2012CB821800), the Natural Science Foundation of China (No. 11473007, 11373038, 11590782) and Strategic Priority Research Program (No. XDB09000000) from Chinese Academy of Sciences.}
}

 \volnopage{ {\bf 2018} Vol.\ {\bf 2} No. {\bf XX}, 000--000}
   \setcounter{page}{1}

   \author{J.S. Zhang
      \inst{1}
   \and D. Li
     \inst{2, 3}
   \and J.Z. Wang
      \inst{4}
  \and  Q.F. Zhu
      \inst{5}
  \and J. Li
      \inst{4}
     }
   \institute{Center for Astrophysics, Guangzhou University, Guangzhou 510006; {\it jszhang@gzhu.edu.cn} \\
   \and        National Astronomical Observatories, Chinese Academy of Sciences A20 Datun Road, Chaoyang District, Beijing 100012
   \and        Space Science Institute, Boulder, CO 80301, USA
   \and        Shanghai Astronomical Observatory, Chinese Academy of Sciences, Shanghai 200030
   \and        University of Sciences and Technology of China, Chinese Academy of Science, Hefei 230026
\vs \no
   {\small Received [year] [month] [day]; accepted [year] [month] [day] }
}
 \abstract{The Five Hundred Meter Aperture Spherical Radio Telescope (FAST) will make contributions to the study of Galactic and extragalactic masers. The telescope, now finished construction and commissioning in China, has an innovative design that leads to the highest sensitivity of any single dish telescope in the world. FAST's potential for OH megamaser research is discussed, including the sky density of masers detectable in surveys. The scientific impact expected from FAST maser studies is also discussed.
  }

   \authorrunning{J.S. Zhang  et al. }            
   \maketitle

\section{Introduction}           

 The Five-hundred-meter Aperture Spherical radio Telescope (FAST) has been constructed successfully and is being tested now. With respect to
 existing single dish radio telescopes (e.g., Arecibo 305\,m, GBT 100\,m, Effelsberg 100\,m, five $\sim$70m ones), FAST has many advantages, namely, much better
 sensitivity due to its unprecedented large collecting area and radio quiet environment due to its unique location.

 The frequency range of FAST is from 70\,MHz to 3\,GHz, within which a main goal of FAST should be OH maser studies. Although the CH$_3$OH maser (J$_K$=1$_1$, $\nu$: 834\,MHz) is another candidate for FAST targets, the only detection of this line was more than 40 years ago (Ball et al. 1970). OH masers ($^2$$\Pi$$_{3/2}$ $J=3/2$, $\nu$: 1612, 1665, 1667 and 1720\,MHz) have been studied for 50 years or so, since their discovery in the 1960s (Weaver et al. 1965). Presently more than 3000 OH masers have been detected in our Galaxy. Galactic OH masers were mostly found in dense molecular gas with star formation (Interstellar masers) and in the molecular circumstellar envelopes of evolved giant and super-giant stars (circumstellar masers, Reid 2002). In addition, one type of OH maser with only 1720\,MHz emission (the other three lines are weak or absent) was detected toward supernova remnants (SNRs), which is  believed to be an unambiguous indicator of SNR shock interactions with molecular clouds (e.g., Frail et al. 1996, Hewitt et al. 2008).

 With improved sensitivity and spectral resolution, OH maser emission was also detected in external galaxies (e.g., Whiteoak \& Gardner 1974, Baan et al. 1982). To date, OH maser emission has been reported from 119 galaxies (e.g., Darling \& Giovanelli 2000, Chen et al. 2007, Fernandez et al. 2010, Willett et al. 2011, Willett 2012, Zhang et al. 2014). Most of them (106/119) have isotropic luminosity larger than 10\,$L_{\odot}$ (also known as ''megamasers''), i.e., a million times more luminous than typical Galactic OH masers. OH megamaser (hereafter OHM) emission is mostly detected in the main line of 1667\,MHz, while the 1665\,MHz main line is weak or absent. The most luminous and distant 1667\,MHz OHM is IRAS\,14070+0525 ($z\sim0.265$, Baan et al. 1992). The satellite transitions at 1612\,MHz and/or 1720\,MHz were only detected in a few nearby galaxies (e.g., McBride \& Heiles 2013). One gravitational lens source, PMN J0134-0931, has been detected at 1720 MHz at z$\sim$0.765 ( Kanekar et al. 2005).

 OH megamaser emission normally has different properties from those of Galactic OH masers. The OHM linewidth (individual components: $>$10\,km\,s$^{-1}$, and total linewidths: $\sim$100-1000\,km\,s$^{-1}$) is at least one order of magnitude broader than that of Galactic OH masers (typically narrower than 1\,km\,s$^{-1}$, Lockett \& Elitzur 2008). OHM emission (mainly at 1667\,MHz) is often unpolarized, while the emission of Galactic masers (mainly at 1665\,MHz) is polarized. Different properties of Galactic OH masers and OHMs should reflect differences in the environment in which the masing occurs and in the excitation mechanism (McBride et al. 2013).
 The hosts of OH megamasers are mostly luminous infrared galaxies (LIRGs, $L_{FIR}$$>$$10^{11}$\,$L_{\odot}$, 102/106), and many are ultra luminous infrared galaxies (ULIRGs, $L_{FIR}$$>$$10^{12}$\,$L_{\odot}$, 35/106). High infrared luminosity galaxies show signs of interaction or merging based on optical imaging observations (Clements et al. 1996). Thus OH megamasers are believed to be good tracers of galaxy interaction or mergers. And the OH megamaser detection rate increases with the far infrared luminosity of the maser host galaxy, up to $\frac{1}{3}$ for ULIRGs (Darling \& Giovanelli 2002; Baan 1991).

 In the following, expectations for FAST observational studies of both Galactic OH masers and extragalactic OH masers are investigated.

\section{Relevant technical specifications}

The illuminated aperture of FAST is of 300\,m diameter (Nan et al. 2011). Assuming an aperture efficiency of 0.65, its effective collecting area is $\sim$46000\,m$^2$, which gives an expected gain of $\frac{A_{eff}}{2760}$ = $16.6$\,$K/Jy$ at L-band. The system temperature of FAST is $\sim$20\,K thanks to its up-to-date receiving system (L-band). Compared with the Arecibo telescope, FAST promises three times better raw sensitivity, which corresponds to about 10 times higher surveying speed (Nan et al. 2011). The site in a deep depression and the feed cabin suspension system enable the FAST reflector a large opening angle, covering a zenith angle up to 40$^{\circ}$ with fully illuminated 300\,m diameter (Nan et al. 2011). At a latitude of N25$^{\circ}$39$^{'}$10$^{''}$, the declination $\delta$ of FAST ranges from about -15 to 65 degrees. Thus the observable solid angle can be estimated: $\omega=\int^{105^{\circ}}_{25^{\circ}}{2\pi\,sin\delta}{\rm d}\delta\simeq2.33\pi$\,$ster.$ ($\sim$24000 square degrees), which is about $\sim$60\% of the whole sky.

The FAST frequency range is covered continuously by nine sets of receivers (Nan et al. 2011, Li et al. 2013). Among those receivers, the L-wide single beam receiver (No.7, 1.2\,GHz--1.8\,GHz) can be used to detect both Galactic and extragalactic OH masers. And observations of OHMs in nearby galaxies can use the 19 beam L narrow Array (No. 8, 1.05-1.45\,GHz). In addition, No. 3 (0.28--0.56\,GHz) and No. 4 (0.56--1.02\,GHz) receivers can be used to detect high redshift OH maser sources (OH megamasers or gigamasers, Zhang et al. 2012).


\section{FAST potential for maser observations}

\subsection{Galactic OH masers}

Surveys, both blind and targeted, facilitate statistical analysis and also result in serendipitous discovery of interesting objects. For Galactic OH masers, blind sky surveys were mostly performed in the 1970s and 1980s with the Parkes 64m, Onsala 25.6m, NRAO 42m, and Effelsberg 100m telescopes (e.g., Johansson et al. 1977; Baud et al. 1979; Caswell et al. 1980, Caswell \& Haynes 1983, Caswell \& Haynes 1987; Sevenster et al. 1997). Meanwhile, targeted Galactic OH maser searches were also performed. Target sources mainly include color-selected IRAS sources (e.g., Eder et al. 1988, Lewis et al. 1990, te Lintel Hekkert 1991; Sevenster et al. 2001), star-forming-regions (e.g., Szymczak \& G\'{e}rard 2004, Wouterloot et al. 1993) and high mass protostellar objects (e.g., Edris et al. 2007).

 FAST has a much bigger collecting area, with a sensitivity at least one order of magnitude better than the telescopes used in blind Galactic surveys
 in the 1970s \& 1980s. To achieve the same sensitivity limit, the FAST surveying speed would be 2 orders of magnitude better. This brings high efficiency for both
 sample selection surveys and Galactic plane blind sky surveys. Many more detections will allow for detailed statistical analysis and kinematic studies of
 interesting sources. The galactic distribution of OH/IR stars (stars with OH maser emission at 1612\,MHz) may provide another good tracer of Galactic structure.

The 1720\,MHz OH masers associated with SNRs were detected in 10\% of galactic supernova remnants (e.g., Wardle \& McDonnell 2012), which are pumped in warm (50$\sim$125\,K) and dense molecular gas environments (Lockett et al. 1999). This type of maser has a typical linewidth of $\sim$1--3\,km\,s$^{-1}$. Observations with five minutes integration give rms noise larger than 20\,mJy in previous surveys (Frail et al. 1996, Hewitt et al. 2008). Assuming a system temperature of $\sim$20\,K and a channel spacing of 0.8\,km\,s$^{-1}$, the rms value of FAST in five minutes is $\sim$1\,mJy. With a sensitivity at least 20 times better, FAST will improve greatly the detection of this type OH masers. It will be of great benefit to understand the interaction between SNRs and adjacent molecular clouds, and further related  questions, such as acceleration of relativistic particles and the structure and physical conditions of the Galactic interstellar medium (Green et al. 1997).

In addition, FAST with much better sensitivity will make it possible in principle to detect Galactic analog OH masers in both massive star formation regions and circumstellar envelopes of evolved stars in very nearby galaxies, e.g., M\,33, M\,31. Comprehensive and comparative analysis of similar OH masers in our Galaxy and nearby galaxies will help to probe various important astrophysical processes related to OH masers. Assuming a typical spectral channel of 4.5\,km\,s$^{-1}$, FAST achieves spectral rms of 0.13\,mJy in one hour of integration time; the three sigma detection threshold is $\sim$0.4\,mJy. Thus OH masers with flux density larger than $\sim$4\,Jy (typical OH maser in our Galaxy) in M\,33 or M\,31 (distance $\sim$800\,kpc) can be detected. VLBI observations of the masers in M\,33 and/or M\,31 would permit accurate measurements of their proper motion and thus reveal the three dimension kinematics of the Local Group ( Brunthaler et al. 2005).

\subsection{OH megamasers}

Surveys of extragalactic OH maser emission have been carried out with the Arecibo 305m, NRAO 91m, JB MkIA 76m, Nancay 300m and Parkes 64m telescopes. Most detections come from the Arecibo telescope and the upgraded Arecibo survey is the most successful one to date (Darling \& Giovanelli 2000, 2001, 2002). A sample of 311 LIRGs or ULIRGs was selected from the IRAS Point Source Catalog Redshift Survey (15,000 IRAS galaxies with $f_{60\,um}$ $>$ 0.6\,Jy over 84\% of the sky, Saunders et al. 2000), with a declination range of 0$^{\circ}$--37$^{\circ}$ (Arecibo sky coverage) and a redshift $z$ range of 0.1--0.45. The lower limit of redshift was set to avoid local radio frequency interference (RFI), while the upper limit was determined by the bandpass of the wide L-band receiver at Arecibo (The actual upper limit is around $z \sim 0.23$ due to the RFI environment). The detection rate was high ($\sim$17\%) and 52 detections doubled the number of known OH megamasers. Recently, one systematic search for high-redshift OH megamasers was carried out with the Green Bank Telescope (Willett 2012). Nine new OH megamasers ($z<0.25$) were detected, with a detection rate of $\sim$7\% from a sample of 121 ULIRGs with $0.09<z<1.5$.

\subsubsection{OH megamaser detectability with FAST:}

By comparison with the upgraded Arecibo OH megamaser survey, we estimate the FAST detectability of OH megamaser emission. Following Darling \& Giovanelli
 (2002), we assume that the integrated line flux can be approximated by 
\begin{equation}\label{1}
\int{S\,{\rm d}V}=f_{peak}\,\frac{\Delta\nu_0}{1+z}=f_{peak}\,\frac{\nu_0\,\Delta\,V_0}{c(1+z)},
\end{equation}
\noindent where $\nu_0$ and $\Delta\nu_0$ are the rest-frame frequency and average frequency width; $\Delta\,V_0$ is the rest-frame velocity width, with a typical value of 150\,km\,s$^{-1}$. Thus the maser luminosity can be estimated as:

\begin{equation}\label{2}
L_{OH}=4\,\pi\,D_{L}^{2}\int{S\,{\rm d}V}=4\,\pi\,D_{L}^{2}\,f_{peak}\,\frac{\nu_0\,\Delta\,V_0}{c(1+z)}
\end{equation}

The luminosity distance $D_{L}$ is derived using the NASA Extragalactic Database (NED) calculator, assuming $\Omega_{M}$ = 0.270, $\Omega_{vac}$ = 0.730, and $H_0$=70\,km\,s$^{-1}$\,Mpc$^{-1}$ (e.g., Spergel et al. 2007). With a 12 minutes integration time and a channel width of $\sim$24.4\,kHz (25\,MHz/1024 channels, which corresponds to velocity spacing of $\sim$4.5\,km\,s$^{-1}$), the typical rms flux density of Arecibo is 0.65\,mJy and the detection threshold at the 3$\sigma$ level is about 2\,mJy (Darling \& Giovanelli 2002). With three times better raw sensitivity and the same integration time and channel width, the rms noise value of FAST is about 0.22\,mJy and 3$\sigma$ level corresponds to 0.66\,mJy ($f_{peak}$). Then, using formula (2), the OH megamaser luminosity for a 3$\sigma$ level FAST detection, with integration time of 12 minutes, can be derived as a function of redshift z and is plotted in Figure\,1. The Arecibo survey OH megamaser detections (black dots) and its sensitivity limit (short black line) are included. In addition, the case of $\sim$2 hours FAST integration time is plotted, which corresponds to 3$\sigma$ detection with $f_{peak}$$\sim$0.2\,mJy. It shows that, in an integration time of 12 minutes, FAST could detect all gigamasers ($L_{OH}>10^{6}\,L_{\odot}$) with redshift z$<$4 and most OH megamasers with $L_{OH}>10^{3}\,L_{\odot}$ out to z$\sim$1, even some out to z$\sim$2. For the case of two hours integration, the majority of OH megamasers with $L_{OH}>10^{2}\,L_{\odot}$ could be detected out to redshift z$\sim$1. All OH megamasers with $L_{OH}>10^{3}\,L_{\odot}$ can be detected out to z$\sim$2, even some out to z$\sim$3 (No. 3 \& 4 receiver). Moreover, as we know, the OH megamaser line has a linewidth of hundreds of km\,s$^{-1}$. Thus tens of km\,s$^{-1}$ spacing is normally good enough for line detection. Assuming a velocity spacing of $\sim$20\,km\,s$^{-1}$ and lower system temperature of $\sim$20\,K, FAST with 12 minutes integration gives rms value of $\sim$2.0\,mK, which corresponds to 0.12\,mJy. In this case, derived minimum detectable OH maser luminosities for 12 minutes and two hours integration are also plotted in Figure\,1, which shows much better expectations.

 \begin{figure}
  \centering\mbox{
  \vspace{-5mm}
  \includegraphics[width=12.0cm]{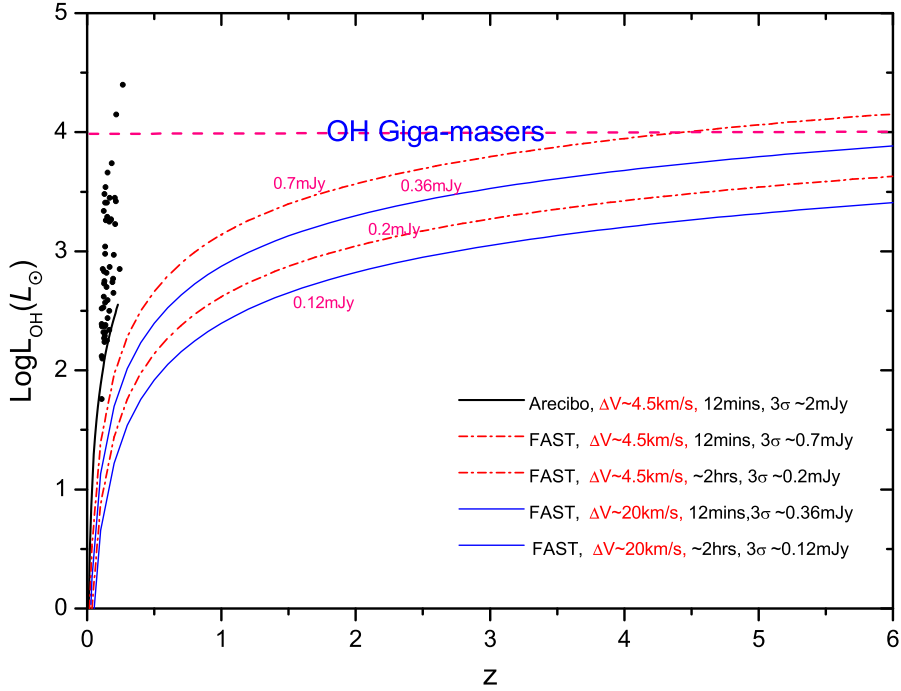}}
  \vspace{-5mm}
  \caption{The minimum detectable OH maser luminosity (3$\sigma$) as a function of redshift, assuming a velocity spacing of $\sim$4.5\,km\,s$^{-1}$ (dash-dotted lines) and $\sim$20\,km\,s$^{-1}$ (solid lines). Cases of 12 minutes and 2 hours integration are presented, respectively. Black dots and short line: detected OH megamsers and sensitivity of the Arecibo survey.}
\end{figure}

 \subsubsection{OH Megamaser sky density:}

The detectable OH megamaser numbers depend mainly on the sky density of OH megamasers, which can be estimated from the OH megamaser luminosity function. The luminosity function from the Arecibo survey is $\phi(L_{OH})=\frac{dN}{dV\,dlog\,L_{OH}}=9.8\times10^{-6}\,L_{OH}^{-0.64}\,Mpc^{-3}dex^{-1}$ (Darling \& Giovanelli 2002), and a new consistent result of $\phi(L_{OH}) = 1.23\times10^{-5}\,L_{OH}^{-0.66}\,Mpc^{-3}dex^{-1}$ was obtained, by combining the Arecibo and GBT new detections (Willett 2012). The number of OH megamasers per solid angle and per frequency range can be determined (Darling \& Giovanelli 2002):

\begin{equation}\label{3}
\frac{dN}{d\Omega\,d\nu}=\int_{log\,L_{OH}^{min}}^{log\,L_{OH}^{max}}{\frac{dV}{d\Omega\,d\nu}\frac{dN}{dV\,dlog\,L_{OH}}\,dlog\,L_{OH}}=
\frac{dV}{d\Omega\,d\nu}\int_{log\,L_{OH}^{min}}^{log\,L_{OH}^{max}}{\phi(L_{OH})\,dlog\,L_{OH}}
\end{equation}

Here, $L_{OH}^{min}$ is the minimum detectable OH luminosity for FAST (12 minutes integration, 3$\sigma$$\sim$0.7\,mJy or 0.36\,mJy, assuming a velocity spacing of $\sim$4.5\,km\,s$^{-1}$ or $\sim$20\,km\,s$^{-1}$, respectively, see Sect.3.2.1) which has been estimated using formula (2). $L_{OH}^{max}$ is assumed to be 10$^{4.4}\,L_{\odot}$ from Darling \& Giovanelli (2002), and $\frac{dV}{d\Omega\,d\nu}$ given by Weinberg (1972). Thus the sky density as a function of redshift can be derived using the luminosity function mentioned above as:

 \begin{equation}\label{4}
\frac{dN}{d\Omega\,d\nu}=k\,\frac{z^2}{\sqrt{\Omega_M\,(1+z)^3+\Omega_\Lambda}}\,(L_{min}^{-0.64}-L_{max}^{-0.64})
\end{equation}
Here, $k=0.0778\,L_{\odot}^{-1}\,MHz^{-1}\,deg^{-2}$.

The merging and interaction rate of galaxies was larger in the past with the rate rising in proportion to (1+z)$^{m}$ (different m values in the literature are $\sim$1--8, e.g., Kim \& Sanders 1998, Briggs 1998, Le Fevre et al. 2000). So the sky density of OH megamasers should also increase with redshift with a factor of (1+z)$^{m}$, under conditions that OH megamasers are good tracers of galaxy merging. Assuming a turnover of evolution at z=2.5 (Briggs 1998, i.e., the merging rate is proportional to $(1+z)^m$ for $z\leq2.5$, but stays constant for $z>2.5$), then the sky density of OH megamasers:

\begin{numcases}{\frac{dN}{d\Omega\,d\nu}=}
  (1+z)^m\,k\,\frac{z^2}{\sqrt{\Omega_M\,(1+z)^3+\Omega_\Lambda}}\,(L_{min}^{-0.64}-L_{max}^{-0.64}), z\leq2.5 \\
  (1+2.5)^m\,k\,\frac{z^2}{\sqrt{\Omega_M\,(1+z)^3+\Omega_\Lambda}}\,(L_{min}^{-0.64}-L_{max}^{-0.64}), z>2.5
\end{numcases}

From equation 5 and 6, the OH megamaser sky density per square degree per 50\,MHz bandpass is estimated and plotted in Figure\,2, including cases of m=0
(without evolution for the merging rate), m=2 and m=4. The detection rate can be predicted for different cases of merging rate evolution. For example, in the case of m=4 and
and a sensitivity of $\sim$0.36\,mJy (assuming a velocity spacing of $\sim$20\,km\,s$^{-1}$), a few OH megamasers at z$\sim$2 per square degree per 50\,MHz bandwidth would be likely detected by FAST, with an integration time of 12 minutes. Dozens of OH megamasers per square degrees within the redshift range $0.6<z<2$ (No.4 receiver, with a bandwidth of $\sim$500\,MHz) may be detected. Even for the case of m=0 and a sensitivity of $\sim$0.7\,mJy, the number of OHMs with redshift $z<0.5$ is expected to be more than 2000 in the FAST sky area ($\sim$24000 square degrees), which will increase the current number of OHMs by 20 times or more.


However, the RFI limits should be considered for OHM surveys at high redshift, as for the Arecibo OHM survey mentioned in Sect.\,3.2. A radio quiet zone of 5\,km around the FAST site has been approved to forbid placement of any new radio transmitters. A 5-10\,km radio controlled zone is also established to require coordination with the FAST project for any new radio transmitters. Within the quiet zone, there is no cell phone base station and we expect the main challenge would be to mitigate our self-generated interference. The L-band RFI environment of the FAST site has been excellent in various tests, some of which were done for SKA site selection. We expect that RFI will not limit our capability of detecting OH mega-masers to higher redshifts except for certain satellite and airport radar bands around 800\,MHz (z$\sim$1).

 \begin{figure}
  \centering\mbox{
  \vspace{-5mm}
  \includegraphics[width=12.0cm]{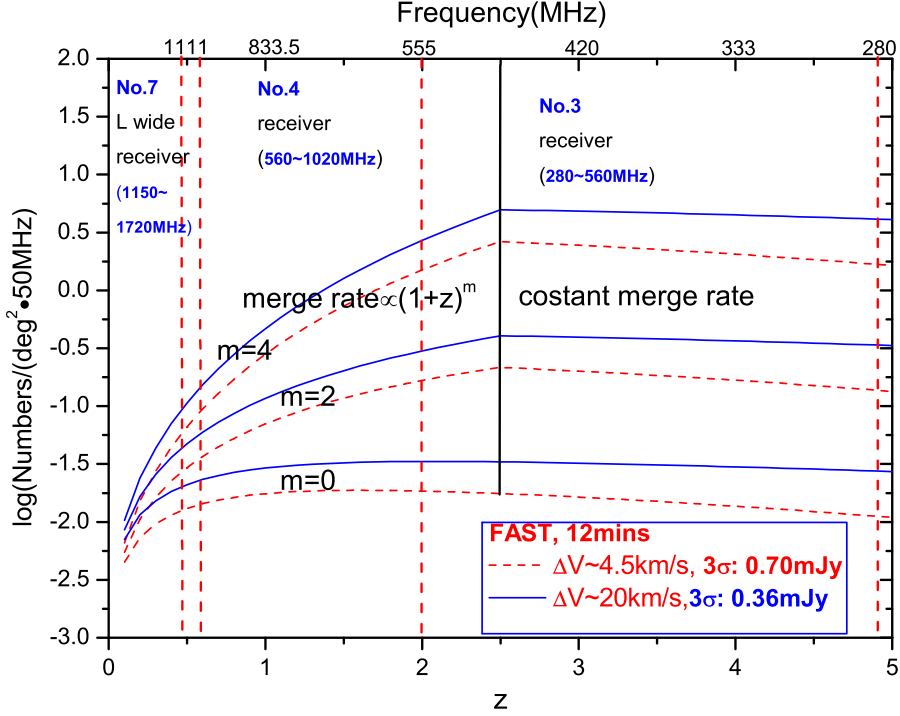}}
  \vspace{-5mm}
  \caption{Detection rate of OH megmasers as a function of redshift (or frequency) with FAST 12 minutes integration (dash lines: 3$\sigma$$\sim$0.7\,mJy, and solid lines: 3$\sigma$$\sim$0.36\,mJy): m=0 for constant merging rate; m=2, 4 for merging rate proportional to $(1+z)^2$ and $(1+z)^4$. The solid vertical line indicates the turnover redshift in merger rate (z=2.5). The frequency ranges of three FAST receivers are also indicated in dashed lines.}
\end{figure}

 \subsubsection{OH megamaser related science}

H$_{2}$O disk megamasers, with maser spots in a rotating molecular accretion disk around the nuclear engine ($\leq$1\,pc), have become a promising tool to probe the innermost region of AGNs (e.g., Morganti et al. 2004) and have important potential for studying cosmology, such as determining accurately the Hubble constant H$_0$ and constraining the equation of state for the elusive dark energy (Braatz et al. 2009). Unlike H$_2$O megamasers, OH megamaser spots are located normally on larger scales ($\sim$100-pc, Lo 2005) from the nucleus within the host galaxy, which may not be good tracers of the circumnuclear accretion disk and further related cosmology. However, as mentioned above, OH megamasers are believed to be good tracers of galaxy interaction or mergers. So OH megamasers could be used to galaxy-merging-related fields, such as the merging history of galaxy merging, the star formation history of the universe, gravitational waves. Given that the merging rate of galaxies is proportion to (1+z)$^{m}$, OHM surveys in the range 400 to 1000\,MHz may provide an independent measure of the merging rate by determining the density of merging galaxies as a function of redshift (e.g., Briggs 1998). And OH megamaser emission is suggested to be physically related to intense starburst activity from galaxy merging, which was supported by existing interferometric studies of several OHM galaxies (e.g., Lo 2005) and statistical analysis of the relation $L_{OH} \propto L_{FIR}^{1.2\pm0.1} $(e.g., Darling \& Giovanelli 2002). Thus detected OH megamasers at different distances should be valuable for understanding the star formation history. In addition, galaxy mergers may produce binary supermassive black holes that in turn produce bursts of gravitational waves (e.g., Jaffe \& Backer 2003).

 \section{Summary}

We present the potential for FAST detections of Galactic masers and extragalactic masers. The innovative design and superior sensitivity of FAST could have significant impact on fields related to masers, particulary Hydroxyl megamasers.

1) FAST with high sensitivity will bring high efficiency in Galactic OH maser surveys and improve detection of 1720\,MHz OH masers related to the interaction between SNRs and adjacent molecular clouds. FAST makes it possible to detect Galactic analogue OH masers in very nearby galaxies, within an integration time of hours (e.g., M\,31, M\,33).

2) Within a reasonable integration time, FAST can detect the majority of OH megamasers with $L_{OH}>10^{2}\,L_{\odot}$ out to redshift z$\sim$1 and all OH megamasers with $L_{OH}>10^{3}\,L_{\odot}$ out to z$\sim$2, even some out to z$\sim$3.

3) The sky density of detectable OHMs is derived as a function of redshift, based on the luminosity function of OHMs and merging evolution scenario parameterized by (1+z)$^m$. For the case of m=4 and a sensitivity of $\sim$0.36\,mJy (assuming a velocity spacing of $\sim$4.5\,km\,s$^{-1}$), with an integration time of 12 minutes, a few OHMs at z$\sim$2 and dozens of OH megamasers within the redshift range $0.6<z<2$ per square degree per 50\,MHz bandwidth would be likely detected by the coming FAST. Even for the case of a constant merging rate (m=0) and a sensitivity of $\sim$0.7\,mJy, the number of detected OHMs with redshift $z<0.5$ could be expected to increase 20 times or more, within the FAST sky coverage ($\sim$24000 square degrees).

\begin{acknowledgements}
This work is supported by the China Ministry of Science and Technology under the State Key Development Program for Basic Research (2012CB821800) and the Natural Science Foundation of China (No. 11473007, 11373038, 11590782). It is also supported by Strategic Priority Research Program (No. XDB09000000) from the Chinese Academy of Sciences. We made use of The NASA Astrophysics Data System Bibliographic Services (ADS) and also the NASA/IPAC Extragalactic Database (NED), which is operated by the Jet Propulsion Laboratory.
\end{acknowledgements}




\end{document}